\documentclass[pra,superscriptaddress,showpacs,twocolumn,nofootinbib,floatfix,amsmath,amssymb]{revtex4-1}
\usepackage{mathtools}
\usepackage[usenames,dvipsnames]{xcolor}
\usepackage{color}
\usepackage{cancel}
\usepackage{braket}
\usepackage{bm}
\usepackage{graphicx}%
\usepackage{hyperref}
\def\op#1{\hat{#1}}

\makeatletter
\@ifundefined{textcolor}{}
{%
 \definecolor{BLACK}{gray}{0}
 \definecolor{WHITE}{gray}{1}
 \definecolor{RED}{rgb}{1,0,0}
 \definecolor{GREEN}{rgb}{0,.4,0}
 \definecolor{BLUE}{rgb}{0,0,1}
 \definecolor{CYAN}{cmyk}{1,0,0,0}
 \definecolor{MAGENTA}{cmyk}{0,1,0,0}
 \definecolor{YELLOW}{cmyk}{.2,.4,1,0}
 }
\makeatother

\begin{document}

\title{Quantum Seismology}
\author{Eric G. Brown}
\affiliation{Department of Physics \& Astronomy, University of Waterloo, Waterloo, Ontario, N2L 3G1, Canada}
\author{William Donnelly}
\affiliation{Department of Applied Mathematics, University of Waterloo, Waterloo, Ontario, N2L 3G1, Canada}
\author{Achim Kempf}%
\affiliation{Department of Applied Mathematics, University of Waterloo, Waterloo, Ontario, N2L 3G1, Canada}
\affiliation{Institute for Quantum Computing, University of Waterloo, Waterloo, Ontario, N2L 3G1, Canada}
\affiliation{Perimeter Institute for Theoretical Physics, 31 Caroline St N, Waterloo, Ontario, N2L 2Y5, Canada}
\author{Robert~B.~Mann}%
\affiliation{Perimeter Institute for Theoretical Physics, 31 Caroline St N, Waterloo, Ontario, N2L 2Y5, Canada}
\affiliation{Department of Physics \& Astronomy, University of Waterloo, Waterloo, Ontario, N2L 3G1, Canada}
\author{Eduardo Mart\'in-Mart\'inez}
\affiliation{Department of Applied Mathematics, University of Waterloo, Waterloo, Ontario, N2L 3G1, Canada}
\affiliation{Institute for Quantum Computing, University of Waterloo, Waterloo, Ontario, N2L 3G1, Canada}
\affiliation{Perimeter Institute for Theoretical Physics, 31 Caroline St N, Waterloo, Ontario, N2L 2Y5, Canada}
\author{Nicolas C. Menicucci}%
\affiliation{School of Physics, The University of Sydney, Sydney, NSW 2006, Australia}

\begin{abstract}
We propose a quantum mechanical method of detecting weak vibrational disturbances inspired by the protocol of entanglement farming. We consider a setup where pairs of atoms in their ground state are successively sent through an optical cavity. It is known that in this way it is possible to drive that cavity toward a stable fixed-point state. Here we study how that fixed-point state depends on the time interval between pairs of atoms and on the distance between the cavity's mirrors. Taking advantage of an extremely precise resonance effect, we find that there are special values of these parameters where the fixed-point state is highly sensitive to perturbations, even harmonic vibrations with frequencies several orders of magnitude below the cavity's natural frequency. We propose that this sensitivity may be useful for high precision metrology.
\end{abstract}
\maketitle
 
\section{Introduction}

 The quantum vacuum is a highly nontrivial state that is also a potential resource for the communication and processing of quantum information.  
In particular, the vacuum state of a quantum field is non-separable when considered with respect to spacelike-separated localized regions \cite{Sorkin1983}. 
This entanglement can be swapped from the vacuum to a pair of spacelike-separated `particle detectors' (such as atoms), which interact with the field locally \cite{Reznik2003}.
Further work showed that this entanglement can, in principle, be used to violate Bell inequalities \cite{Reznik2005}. 
The amount of entanglement depends on properties of the field, of the background spacetime \cite{VerSteeg2009,Nambu:2013gx} and  atom trajectories~\cite{Saltoninprep}. 

 This phenomenon has come to be known as `entanglement harvesting'~\cite{MartinMartinez2012}, and the results reported in \cite{VerSteeg2009,Saltoninprep} already suggest that entanglement harvesting can be a useful tool for metrology.   While the preceding work shows that the vacuum of a quantum field possesses entanglement that can be harvested in principle, the amount that can be extracted tends to be exceedingly small unless the  atoms are very close together. Furthermore, if we wish to employ this entanglement for any useful purpose, we need to know how repeatedly extracting entanglement affects the background resource (i.e.,~the field state). We can solve both of these problems by moving from a hunter-gatherer approach to an agricultural one: we can use \emph{entanglement farming} instead.

First introduced in Ref.~\cite{farming}, entanglement farming involves successively sending pairs of `particle detectors' (such as atoms, ions, molecules, etc.)\ transversely through an optical cavity, all initialized in their ground states. As each pair of atoms\footnote{Henceforth,  we use the term `atom' for   the generic system interacting with the cavity field.} traverses the cavity, the state of the cavity field is slightly modified. As pair after pair traverses the cavity, the field approaches a fixed-point state through a non-perturbative and non-thermal process, as was shown in \cite{farming}. When the fixed point is reached, every pair of atoms emerges from the cavity in the same state, which is generically entangled. Due to the stability of the fixed-point state, this protocol provides a potentially useful method for producing a stream of reliably entangled pairs. Additionally, this protocol was proven to be robust to variation of the parameters and, most importantly, almost entirely independent of what the initial state of the field---in particular, not requiring it to be the vacuum state for the fixed point to be quickly reached. Since this process yields a sustainable source of harvested entanglement, the term `farming' appears appropriate.

 Entanglement farming depends on the \text{(meta-)stable} fixed point of the cavity that is produced by successively passing  pairs of atoms through the cavity. This fixed point can be calculated using non-perturbative continuous-variable methods \cite{Eric} and it was found to be generally stable to small changes in the parameters of the setup (e.g.,~positions, time of flight,  energy gap, cavity length, etc.)~\cite{farming}.
Here we will show that this robustness breaks down dramatically when the frequency at which atoms traverse the cavity is at resonance with a multiple of the cavity's fundamental frequency. 
Concretely, we can tune the parameters (including the waiting time between pairs of atoms) so that the steady state is highly sensitive to  changes in these other parameters. This finding opens up opportunities to use this setup to detect small parameter changes with very high sensitivity.     In what follows, we choose this free parameter to be the cavity length---i.e.,~we study the sensitivity of entanglement farming to small deviations in the length of the cavity, which makes our setup sensitive to vibrations. 

A technical aspect of our approach relates to the fact that time-dependent boundary conditions introduce nontrivial effects on quantum fields, such as particle creation by moving mirrors, sometimes called the dynamical Casimir effect~\cite{Davies1976,dynCas}. 
We will restrict our analysis to settings in which these effects are negligible. It will turn out that the sensitivity of our setup to the parameters of interest remains significant even in this adiabatic regime.  The sensitivity furthermore remains even when the frequency of vibration is several orders of magnitude below the fundamental optical frequency of the cavity,  making this potentially a very sensitive apparatus for detecting mechanical perturbations of optical cavities---a kind of \emph{quantum seismograph}.

\section{Setup}

The  setup of the quantum seismograph is based on the farming scheme \cite{farming}. Consider a scalar quantum field in a cavity of fixed length $L$ with Dirichlet boundary conditions. The general solution of the field equations can be written as a mode expansion of the form
\begin{equation}\label{eq:modeexp}
\phi(x,t) = \sum \varphi_n(t) \sin (k_n x).
\end{equation}
In terms of these variables the field is described as a system of uncoupled harmonic oscillators with wavenumbers $k_n$ and frequencies $\omega_n$, where $k_n = n \pi / L = \omega_n / c$. For convenience, we will use natural units ($c=1$) for the rest of the paper.

We consider pairs of atoms in their ground state that cross the cavity sequentially, each of which couples to the field in the cavity via the interaction-picture Unruh-DeWitt interaction Hamiltonian \cite{DeWitt,Crispino} %
\begin{equation}
\label{eq:UdW}
H_I=\lambda \chi(\tau) \mu(\tau) \hat \phi[x(\tau),t(\tau)], 
\end{equation}
where $\lambda$ is an overall coupling constant (not necessarily small); $\chi(\tau)$ is a switching function that encodes the time dependence of the coupling strength; $\mu(\tau)= e^{-i\Omega\tau} b+e^{i\Omega\tau} b^\dagger$ is the interaction-picture monopole-moment operator of the atom, which we model as a first-quantized harmonic oscillator with frequency $\Omega$ (thus with free Hamiltonian $H_0^{(\text{d})}=\Omega b^\dagger b $); $x(\tau)$ is the worldline of the atom parametrized in terms of its proper time; and $\hat\phi[x(\tau),t(\tau)]$ is the interaction-picture quantized field operator, which we expand in terms of modes as prescribed in \eqref{eq:modeexp}.   In our current scenario the atoms will remain stationary during the course of interaction with the field, $x(\tau)=\text{const}$. This means that the proper time of each   is equal to the global field time parameter $\tau=t$.  

Equation~\eqref{eq:UdW} is a common simplified model for  light-matter interaction~\cite{DeWitt,Birrell,Scullybook} that captures all its essential features when no orbital angular-momentum transitions are considered~\cite{Wavepackets,Alvaro}. We are going to use recently developed tools~\cite{Eric,Ivette},   which have been applied to the study of a number of relativistic quantum phenomena \cite{Brenna13,BrownTherm},  to carry out a non-perturbative analysis of the atom-field dynamics. Following Ref.~\cite{farming}, we will make use of the continuous-variable non-perturbative formalism reported in Ref.~\cite{Eric}.

In this context we analyze the dynamics of two atoms of equal energy gap~$\Omega$, initialized in their ground state and which only interact with the field for a finite amount of time~$T$. After this time, the two original atoms are removed, and a fresh pair is set to interact with the field in the cavity, again for a time $T$. We repeat the whole process iteratively, eventually reaching a fixed point~\cite{farming} and recovering pairs of entangled atoms.  In the same fashion as in \cite{farming}, the physical implementation of such setting  consists of beams of atoms traversing the cavity in a direction perpendicular to the quantization direction $x$, in a similar way as in \cite{Haroche}. 

The fact that this fixed point does not depend on the initial state of the field considerably reduces the challenge of experimentally implementing these settings. For instance, quantum optics provides a way to physically implement this repeated sequence of pairs of atoms interacting with the field: We can shoot pairs of atoms transversely through the cavity in a scheme similar to that of Ref.~\cite{Haroche}. In this case, the switching function will be given by the spatial profile of the cavity's transverse spatial modes and by the speed of these atoms through the cavity. Alternatively, one could think of superconducting-circuit schemes \cite{ultrastrongOriginal}, where it is possible to implement a controllable ultra-strong switchable coupling \cite{ultrastrong}.

Given the Dirichlet boundary conditions linking length and frequency scales, only time (or alternatively length) units are free to be chosen: We let the fundamental frequency of the cavity $\omega = \pi/L$ carry the relevant units for the physical system in question. Hence, all the other quantities of the simulation should now be interpreted relative to this fundamental frequency. For instance, a cavity whose fundamental frequency is 10~GHz (microwave cavity) corresponds to a length of roughly $L\approx 3$~cm. If the frequency is 500~THz then $L$ would be roughly $\approx 600$~nm.

\section{Sensitivity of the fixed point to time delays}
  Let us now introduce a delay of some duration~$\Delta t$ between the exit of one pair of atoms and the entry of the next pair. During this delay, the field will undergo free evolution. A natural question arises: To what extent does the fixed-point state depend on~$\Delta t$? We find that the introduction of such a delay typically does not strongly affect the steady state. However, we show below that for delays in the vicinity of particular isolated critical values of $\Delta t$, the steady state can vary greatly with very small changes in this delay.

\begin{figure}[t]
	\centering
        \includegraphics[width=0.5\textwidth]{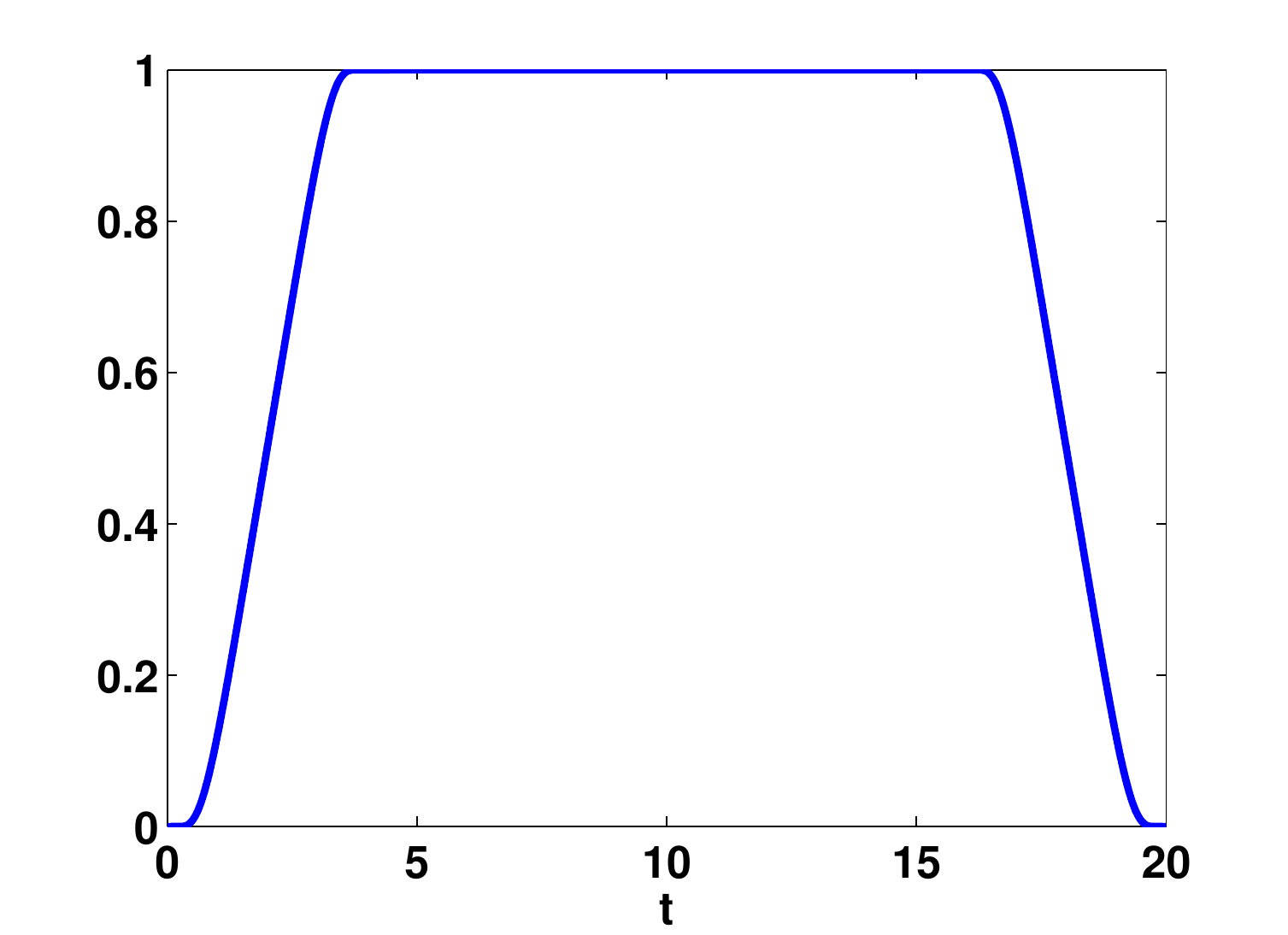}
	\caption{The function $\chi(t)$ with $T=20$ with $\delta=0.2 T$.}
        \label{smoothFunc1}
\end{figure}

To eliminate any possible spurious effects coming from an ill-defined sudden switching \cite{Louko:2007mu},  we will ramp up the strength of the interaction between the atoms and the cavity modes with the following smooth~($C^\infty$), compactly supported switching function \cite{echo}:
\begin{align}
\label{eq:chit}
\chi(t) &= 
\begin{cases}
    S\left[ \pi{t}/{\delta} \right]      & 0\le t < \delta,\\
    1                                   & t\in[ \delta, T -  \delta), \\
    S\left[ \pi({T-t})/{\delta}\right]  & T-\delta \le t \le T, \\
    0                                   & \text{elsewhere},
\end{cases}
\end{align}
where $S(x) = [1-\tanh(\cot x)]/2$. This function smoothly switches from $0$ to the full coupling strength $\lambda$ and back to 0, where $\delta$ is the switching time-scale. In Fig.~\ref{smoothFunc1} we plot an example of this function that we will use in our scenario, with $\delta=0.2T$ and $T=20$. 

We  shall not  solve for the dependence of the fixed point on $\Delta t$ analytically. Instead, we will uncover this dependence numerically. To this end, let us employ the following system parameters. The coupling constant is $\lambda=0.01$.  The boundaries of the cavity are located at $x=0$ and $x=L_0$ and the two  atoms  are located at $x_1=L_0/3$ and $x_2=2L_0/3$ (such that the distance between them is $L_0/3$).   We choose the frequency of the atoms to be resonant with the fundamental mode of the cavity: $\Omega=\pi/L_0$. The time of interaction for each cycle (i.e.,~how long each pair of atoms spend in the cavity) is $T= 2.5 L_0$. Note that this is well beyond the light-crossing  time~$L_0/3$ between the atoms, as required for the farming procedure to work~\cite{farming}.  The choice that we have made here for $L_0$ is entirely arbitrary. As will be discussed later, we can scale down the cavity length to that of an optical cavity or cavity QED setup and, by similarly scaling the other dimensionful quantities, we can obtain exactly the same results. Indeed, we will discuss how the results we obtain here with the above parameters are equivalent to what can currently be achieved within cavity QED systems.

\begin{figure*}[t]
	\centering
                 \includegraphics[width=0.48\textwidth]{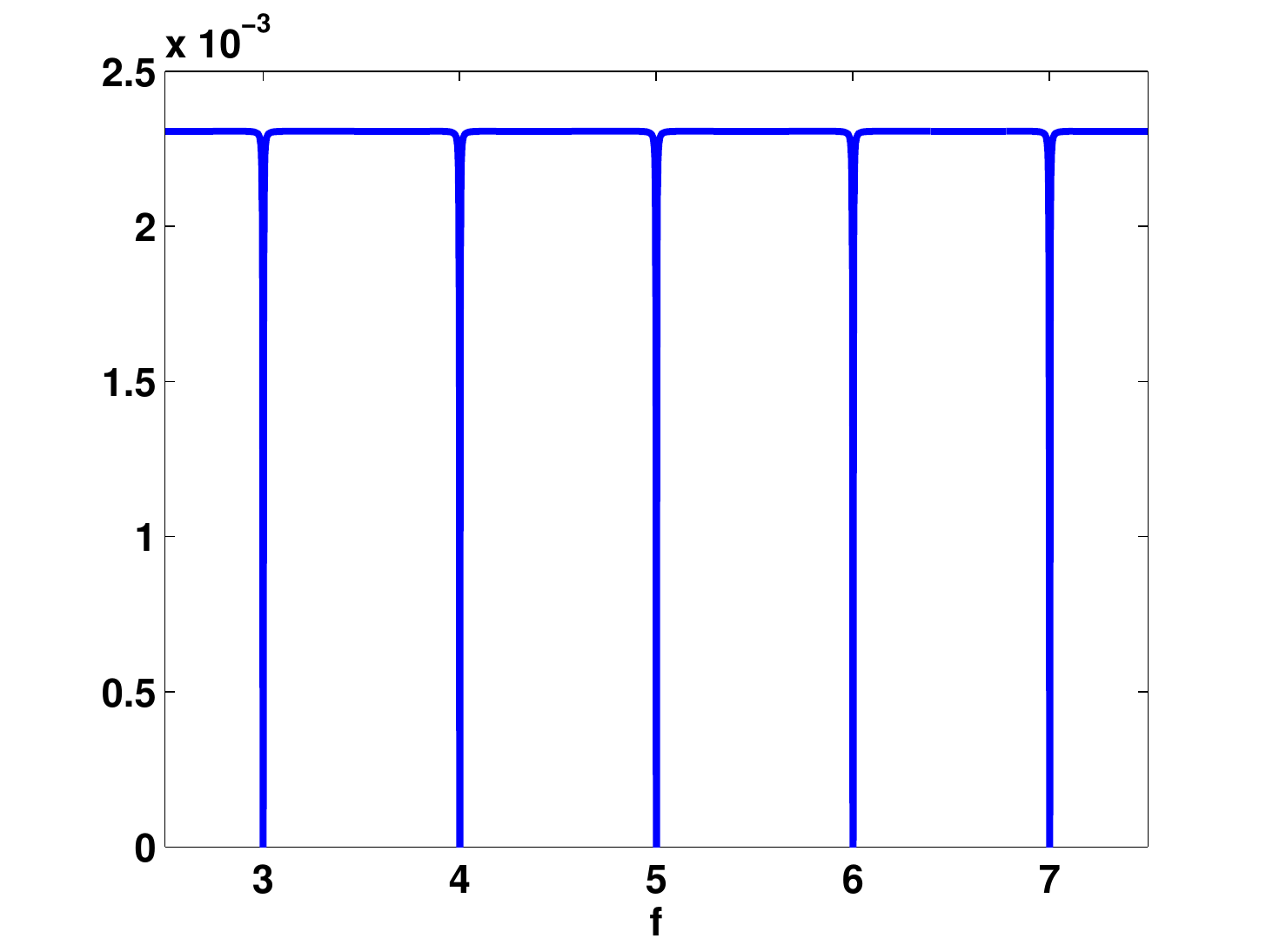}
                 \includegraphics[width=0.48\textwidth]{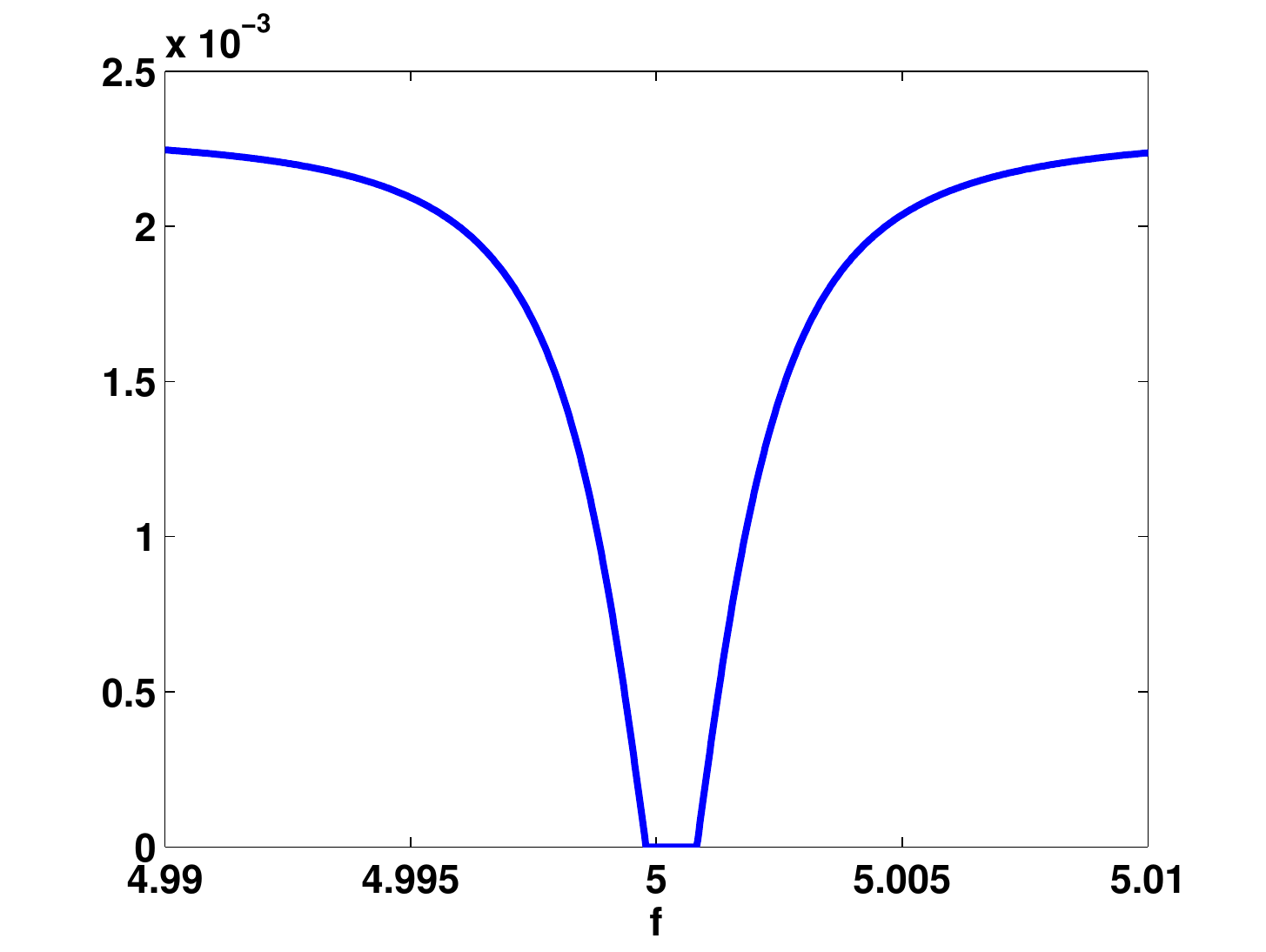}
	\caption{The steady state logarithmic negativity as a function of $f=(T+\Delta t)/L_0$, where $\Delta t$ is the varied parameter. The other parameters are %
	$\Omega=\pi/L_0$ (resonant with the fundamental mode), $T=2.5 L_0$, and with the switching parameter $\delta=0.2 T$. On the left we see that the steady state entanglement is nearly constant as the delay time is changed, except for very well-defined points in which it drops drastically. On the right we zoom in on one of the valleys, and we see that in fact the steady-state entanglement is zero within a small window.}
        \label{valleys}
\end{figure*}

  We plot in Fig.~\ref{valleys} the logarithmic negativity of the state of a pair of atoms once the fixed-point state is reached, as a function of the time between successive pairs of atoms traversing the cavity, in units of the light crossing time of the cavity, $f=(T+\Delta t)/L_0$. 
As Fig.~\ref{valleys} shows, there are remarkably sharp valleys  at integer values of $f$.

The presence of these valleys suggests an interpretation as a resonance effect.
By iterating our protocol, we introduce a periodic time-dependent perturbation to the Hamiltonian of strength $\lambda$ whose frequency is at resonance with the modes of the cavity precisely when $f$ is an integer. 
When $f$ is close to an integer, this perturbation induces transitions between levels of the system, which in our setup corresponds to emission and absorption of field quanta by the atoms.
The width of such a resonance in frequency space scales as $\lambda^2$ with the coupling constant $\lambda$, which explains the sharpness of the valleys.
Note that although our perturbation is not harmonic in time (its shape is given by the switching function in Fig.~\ref{smoothFunc1}), our results suggests it is only the periodic nature of the perturbation that is important.
Indeed, we have investigated how the valleys depend on the switching function, and we found almost identical valleys even for completely sharp switching. While this suggests that the simplifying assumption of sharp switching could be used instead of smooth (as in Ref.~\cite{farming}) without creating artifacts, we nevertheless continue to use the smooth function, Eq.~\eqref{eq:chit}, to ensure maximum confidence in our results. 

The sharpness of these valleys---i.e.,~the extremely strong sensitivity of the fixed-point state---suggests applications to metrology. The idea is to prepare an entanglement farming system with initial parameters such that the steady state is within one of these valleys, preferably at the steepest point of one of the valley's walls. Then, even a weak disturbance (for example, a tiny change in the length of the optical cavity) may displace the system out of this sharp valley and cause a significant change to our readout, yielding a strong signal. As we will show, this does occur.  Importantly, not only do we receive a remarkably strong signal in the entanglement between atoms but also directly in more measurable quantities such as the quadrature correlation functions. 

\section{Cavities with time dependent length}
\label{sec:cavitylength}

 Let us now suppose that the cavity is disturbed, say by a mechanical wave of some kind, so that the cavity's proper length becomes time-dependent. 
When a mechanical wave deforms the cavity we will make the  assumption that the atoms  keep their positions constant relative to the instantaneous cavity length, e.g. $x_1=L(t)/3$ and $x_2=2L(t)/3$. This is a conservative approach that will yield a lower bound on the setting sensitivity: If the atoms moved  relative to the cavity (in the longitudinal direction), they would likely feel an even larger disturbance given the variation of their effective coupling strength due to the inhomogeneous spatial profile of the field modes. 

Keeping the relative positions constant, any change we see in the farming output must result solely from a change of field state rather than just a direct change in the coupling strength of the atoms to the field modes, which of course also induces a detectable change in the atoms' dynamics \cite{farming}.

Intuitively, as long as the time-dependence of the cavity length is slow enough, the modes in that cavity should be approximately the same as those for a stationary cavity, except that each mode's frequency now varies in time. We call this assumption (specifically, $\lvert \dot L \rvert \ll 1$) the \emph{adiabatic approximation}.
because it is equivalent to the usual adiabatic approximation $\dot \omega \ll \omega^2$ in terms of the mode frequencies $\omega \sim 1/L$.
We know that when the cavity walls' speed is comparable to that of light ($\dot L \sim 1$), relativistic effects render this naive description inaccurate (see, e.g., the dynamical Casimir effect~\cite{dynCas}). Fortunately, there exists a wide range of wall motion parameters that (a)~are consistent with the adiabatic approximation and (b)~produce an observable disturbance of the entanglement farming process.

In other words, the system of uncoupled harmonic oscillators becomes a system of uncoupled oscillators with time-dependent frequencies $\omega_n(t)$ given by
\begin{equation} \label{adiabatic}
\omega_n(t) = k_n(t)=\frac{n \pi}{L(t)} 
\end{equation}
which in turn means that the Heisenberg-picture free-field Hamiltonian is
\begin{align}
\label{eq:Hheis}
 H &= \sum_n \frac{\omega_n(t)}{2} \left(\pi_n^2(t) + \varphi^2_n(t)\right) +\mathcal{O}[\dot L (t)].
\end{align}
In Appendix~\ref{app1}, we justify the form of this Hamiltonian and our use of the adiabatic approximation for a cavity comprised of one fixed mirror and another undergoing forced oscillations. As a possible extension of this work, Appendix~\ref{app2} provides preliminary calculations for a cavity deformed by the passage of a gravitational wave, producing a similar distortion.In the rest of the paper, we assume the adiabatic approximation in this physical scenario.

\section{Quantum seismograph} 

\subsection{Two-stage evolution}

The idea behind the quantum seismograph is to prepare an entanglement-farming setup in a steady-state configuration with initially fixed cavity length, such that small changes in the length of the cavity, due to temporary vibrations, produce a detectable change in the extracted entanglement and other measurable quantities.

Modelling the evolution of this system involves two stages. The first is to calculate the steady-state response of the system to a variety of (fixed) parameters, including cavity length~$L_0$, interaction time~$T$, and delay between interactions~$\Delta t$, in order to determine which ones should be used as the unperturbed cavity parameters. These parameters are assumed to be constant during this first stage. We model the evolution using the symplectic methods of Ref.~\cite{Eric}. There are two important differences with respect to the setup of Ref.~\cite{farming}: (a)~for each cycle of atoms entering and exiting the cavity, we model the interaction Hamiltonian of Eq.~\eqref{eq:UdW} using the smooth window function~$\chi(t)$ from Eq.~\eqref{eq:chit} (see Fig.~\ref{smoothFunc1}) with overall coupling constant $\lambda=0.01$ and interaction time per cycle $T=2.5 L_0$ (with switching parameter $\delta=0.2 T$); and (b) this interaction is followed by a delay (i.e., free evolution of the field) of duration~$\Delta t$. The fact that, due to~(a), the evolution is no longer piecewise time-independent means that we must numerically integrate the equations of motion~\cite{Eric} instead of simply numerically evaluating their analytic solution as was done in Ref.~\cite{farming}. The system is allowed to reach a steady state, which is assumed to be the unwavering behavior of the system before any vibrations have affected it.

For the second stage of the evolution, we imagine that the system is steadily humming along in its steady state (as above) when suddenly the cavity experiences a vibration, resulting in a
sinusoidal variation in cavity length,
\begin{align}
\label{eq:Loft}
	L(t)=L_0+A\sin(\gamma t) \qquad (t>0),
\end{align}
with amplitude~$A$ and angular frequency~$\gamma$. To keep within the adiabatic approximation (see Sec.~\ref{sec:cavitylength}), we require
\begin{equation}
    \gamma A \ll 1,
\end{equation}
which ensures that the cavity walls' motion is nonrelativistic ($\lvert \dot L \rvert \ll 1$).\footnote{Notice that this allows $A$ to be large as long as $\gamma$ is sufficiently small. In practice, however, if we want to detect weak vibrations, then we expect $A \ll L_0$, as well.} We do not calculate the steady state of the system during this (time-dependent) evolution, since we want to consider the case where the system does not have time to reach the steady state (if such a state exists).  
Instead, we numerically integrate the full dynamics~\cite{Eric} for this stage of the evolution, using the time-dependent free-field Hamiltonian given in Eq.~\eqref{eq:Hheis}, along with the atoms' interaction, Eq.~\eqref{eq:UdW}, and delay as described above. The initial state for the evolution is taken to be the steady-state solution resulting from the previous stage of the evolution.
From our simulations, we extract information about observables associated with each pair of atoms after it exits the cavity.

\subsection{Choice of unperturbed cavity parameters}

In order to maximize the response we get from a change in~$L$, we optimize our choice of unperturbed cavity parameters as follows. We calculate the logarithmic negativity \cite{logneg}  
of the exiting atoms using a variety of constant parameters. In particular, we can prepare the system initially such that its steady state corresponds to a point on a very steep part of one of the ``valleys" seen in Fig.~\ref{valleys}. The idea behind this choice is that small periodic changes in~$L$ should result in movement along this steep ``wall'' of the valley, producing a large change in the extracted entanglement with detectable time dependence. This intuition relies on the tacit assumption that the steady-state plot shown in Fig.~\ref{valleys} is still relevant in stage two of the evolution (i.e., full dynamics, including vibrations in~$L$, Eq.~\eqref{eq:Loft}). We expect this intuition to be valid when a single period of the vibration lasts over many interaction cycles (including both the interaction time~$T$ and the time of free evolution~$\Delta t$)---in other words, if we choose
\begin{equation}
    \gamma (T + \Delta t) \ll 1.
\end{equation}
We must stress, however, that this assumption is not required for numerical stability of the simulation or for validity of the results we calculate. This is because the steady-state calculations (which use fixed~$L_0$) are used only to determine the initial state of the cavity field. The simulation calculates the full evolution of the system starting from this state, as discussed in the previous subsection.

\subsection{Detecting vibrations}

We choose the time of interaction to be $T=2.5 L_0$ and the rest time of the field to  be the same $\Delta t = T$,   which puts us into the valley at $f=5$ as seen in Fig.~\ref{valleys}.  We allow the system to reach its steady state. Then, as seen in the figure, we receive no entanglement from each pair of  atoms that emerges from the cavity. Thus, if nothing disturbs the system a steady stream of unentangled pairs emerges from the cavity.

\begin{figure}[t]
	\centering
        \includegraphics[width=0.5\textwidth]{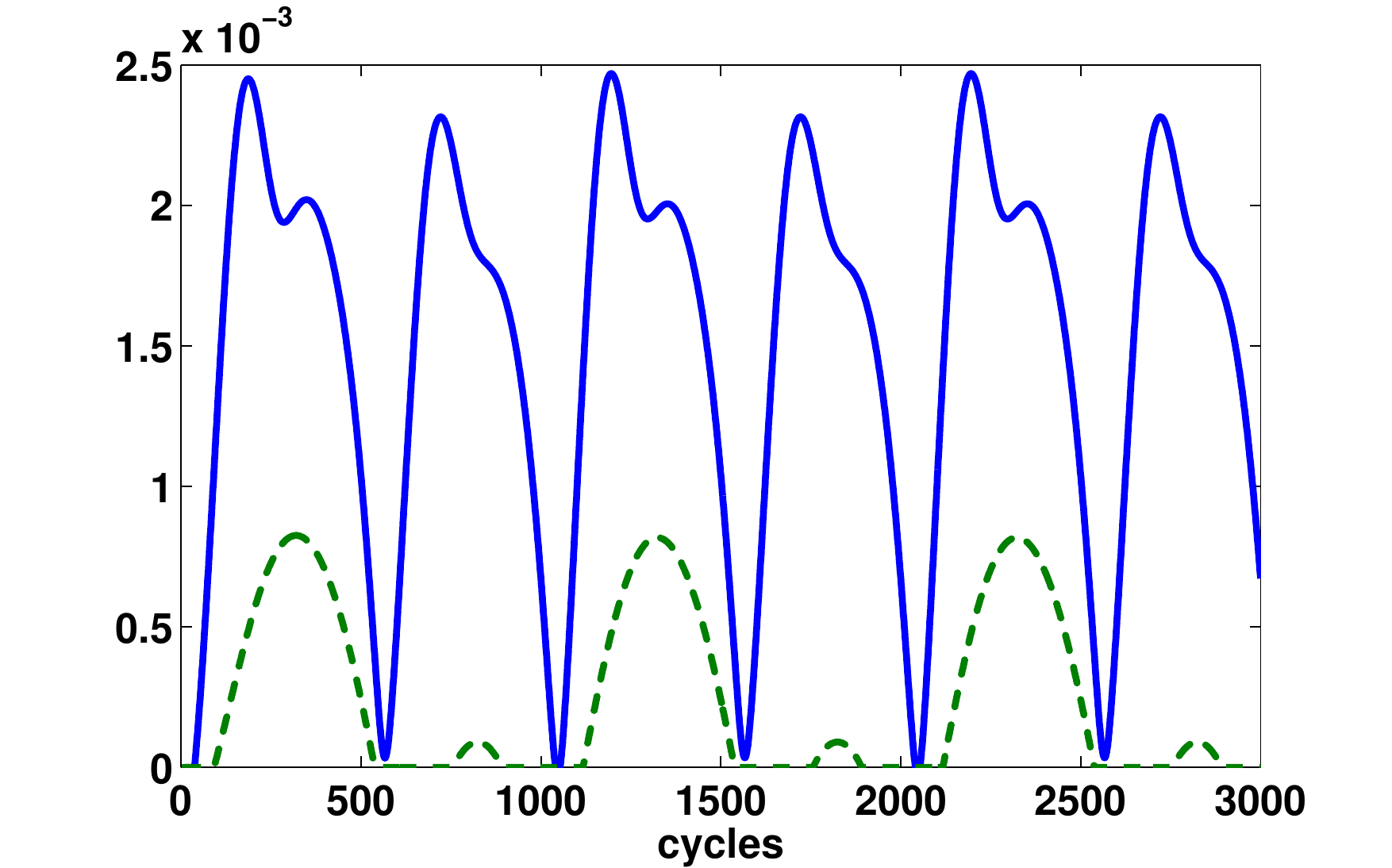}
	\caption{The logarithmic negativity received per cycle during the period of vibration. Here the vibrational period is set to $1000$ times the cycle time (interaction plus rest period totaling a time of $T + \delta t = 2T$) and we track three periods of the vibration. Before the vibration the steady-state entanglement was zero, and the entanglement shown in this plot is due solely to the vibration, despite the frequency being such that $\gamma/\omega_1=4\times 10^{-4}$. The two lines correspond to different vibration magnitudes. The solid (blue) line is such that $A=(1\times 10^{-3}) L_0$ and the dashed (green) line corresponds to $A=(2\times 10^{-4})L_0$.}
        \label{entResponse}
\end{figure}

We now let a vibration occur that is weak in magnitude and of low frequency (as compared with the optical frequency of the cavity). As stated above, by setting $f=5$ we have prepared our system within a precariously thin ``valley". Does this indeed mean that the fixed point is extremely sensitive to even such a non-invasive disturbance? 
Consider an example where the frequencies of the atoms are $\Omega=\pi/L_0$ (resonant with the first mode). 
Considering the logarithmic negativity per cycle of each successive pair of atoms, which is initially zero, we see in  Fig.~\ref{entResponse} that a very significant response is obtained due to the presence of a wave. In this example the frequency of the wave is $\gamma=(4\times 10^{-4})\omega_1$, where $\omega_1$ is the   fundamental frequency   of the cavity, and the amplitudes   corresponding to the two responses shown are $A=(1\times 10^{-3}) L_0$ and $A=(2\times 10^{-4}) L_0$.  While the value of the logarithmic negativity obtained is small, recall that to leading order the logarithmic negativity generated between two atoms interacting with a field goes as $\lambda^2=10^{-4}$.  Here, we are finding that a very weak and low-frequency disturbance can cause the generated entanglement to jump from a zero value to even an order of magnitude \emph{higher} than was generally expected from the harvesting scenario. In this sense, we have found an extremely strong signal. 

  Whereas this variation in the per-pair entanglement could be amplified through entanglement distillation, it is not very convenient experimentally to rely on the amount of produced entanglement to encode the information about the perturbation. While it is interesting to see how the generated entanglement is affected by a mechanical a vibration of the cavity, we should also consider the impact on other, more directly measurable, quantities.    

\begin{figure}[t]
	\centering
        \includegraphics[width=0.5\textwidth]{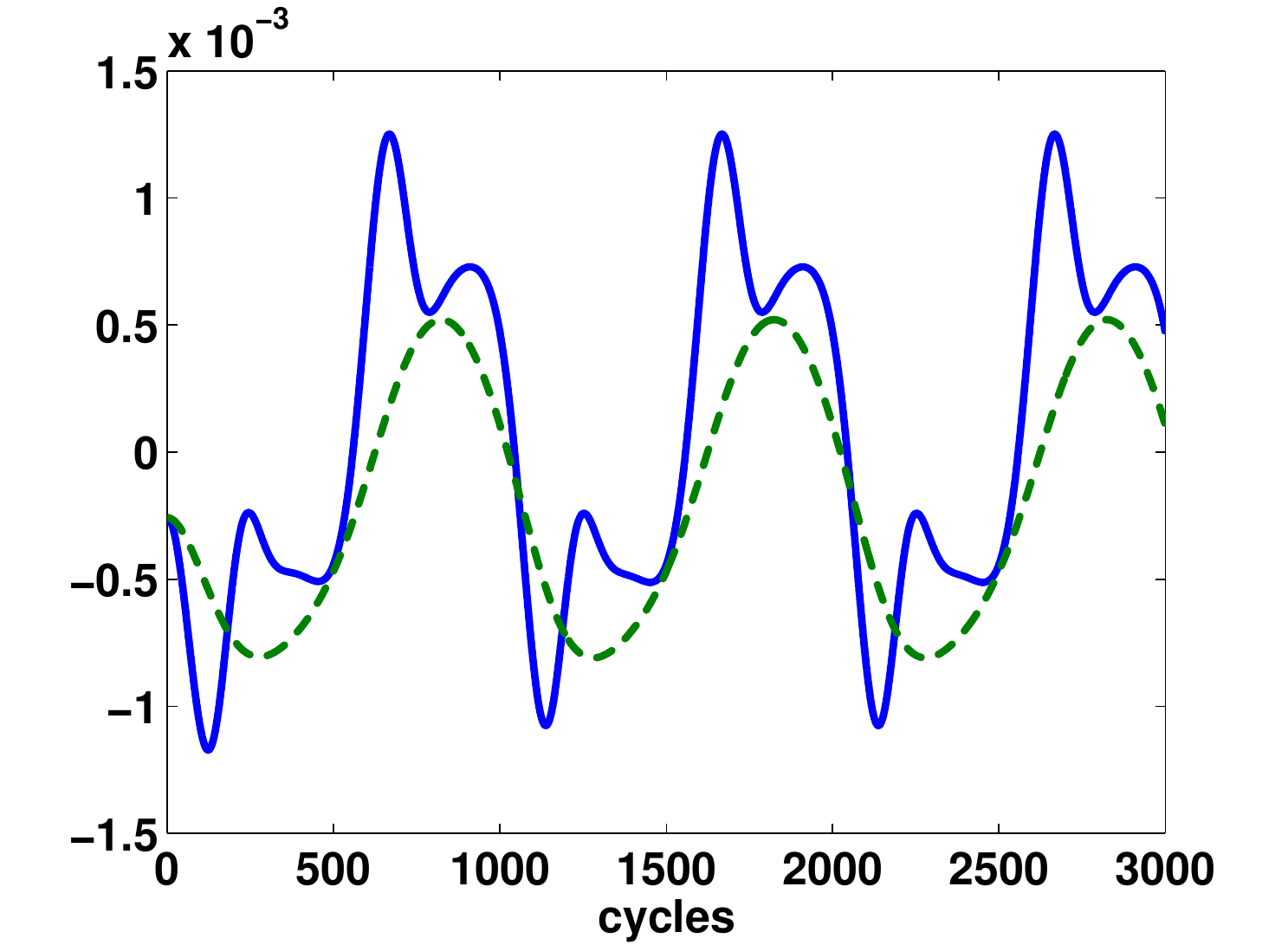}
	\caption{The correlation function $2\braket{\op q_1 \op p_2}$ per cycle during the period of vibration. The specific scenario and parameters are the same as in  Fig.~\ref{entResponse}. The solid (blue) line corresponds to a vibrational amplitude of $A=(1\times 10^{-3}) L_0$ and the dashed (green) line to $A=(2\times 10^{-4})L_0$. Note that we achieve an order-of-magnitude increase in this quantity due to the presence of the wave, as compared to the steady state value (approximately $-0.25 \times 10^{-3}$, as given by the initial value in this plot).}
        \label{corr}
\end{figure}

For example we can consider the entries of the covariance matrix directly. Here we will look at the  observable $\braket{\op q_1 \op p_2+\op p_2 \op q_1}=2\braket{\op q_1 \op p_2}$,  where $\op q_1$ and $\op p_2$ are the internal conjugate position of atom 1 and momentum of atom 2, respectively.  Let us consider the exact same scenario and parameters as those used in Fig.~\ref{entResponse}. The result is displayed in Fig.~\ref{corr}.   Note that we achieve an order-of-magnitude variation from the steady-state value of $2\braket{\op q_1 \op p_2}$ (approximately $-0.25\times 10^{-3}$, as given by the initial value) due to the presence of the wave. 
Although we chose to focus on the $\hat q_1 \hat p_2$ correlator, the other entries of the covariance matrix behave similarly.

\subsection{Frequency response}

\begin{figure}[t]
	\centering
        \includegraphics[width=0.5\textwidth]{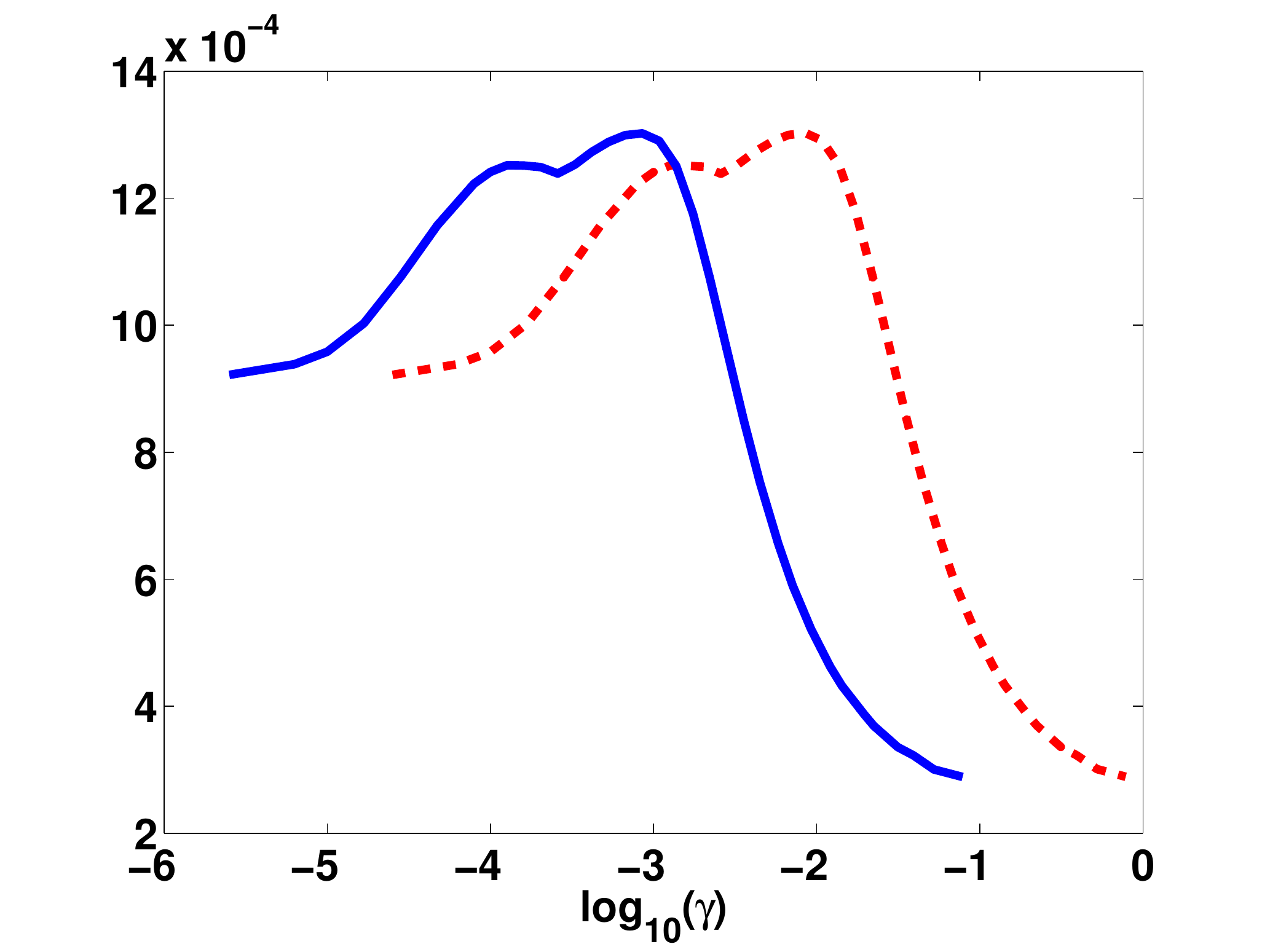}
	\caption{The maximum value of $2\braket{\op q_1 \op p_2}$ achieved over the course of $10$ vibrational periods as a function of the log of the vibrational frequency $\gamma$. The parameters for the solid line (blue) are the same as for Fig.~\ref{corr}: $L_0=8$, $\Omega=\pi/L_0$, $\lambda=0.01$, and $T=20$ with $\Delta t=T$. The dashed line (red) represents the case that all parameters have been scaled by an order of magnitude: $L_0=0.8$, $\Omega=\pi/L_0$, $\lambda=0.1$, and $T=2$ with $\Delta t=T$. In both cases the amplitude of vibration is assumed to scale with the initial length, such that $A=(1\times 10^{-3}) L_0$. We see that by scaling our system in such a way we can achieve sensitivity coverage over a range of vibrational frequencies.}
        \label{freqRes}
\end{figure}

  An important question that must be answered is to what degree our proposed system is sensitive to a range of vibrational frequencies $\gamma$. To this end, we take a vibrational magnitude of $A=(1\times 10^{-3})L_0$ and consider the response due to a range of $\gamma$ spanning over several orders of magnitude.  For each frequency the wave will last for $10$ periods, over however many atom-field interaction cycles are required for this time period. For each we will then take the maximum magnitude of $2\braket{\op q_1 \op p_2}$ achieved over all cycles that occur during the wave. Figure \ref{freqRes} plots this quantity as a function of $\log_{10}\gamma$ for two different sets of parameters, showing that our proposal can be tuned to be sensitive to a wide range of different frequencies.  
  
We observe that for a given set of parameters we obtain a well-defined region of sensitivity, and furthermore by rescaling the parameters of our setup we can tune this to a region of our choosing. This rescaling involves modifying the initial length $L_0$ of the cavity (thus changing the fundamental frequency) and also scaling the other dimensionful quantities accordingly, such that $\lambda L_0$, $\Omega L_0$, $T/L_0$, $\Delta t/T$, and $A/L_0$ remain the same. Such a scaling leaves invariant the dynamics of the system, and the exact same results are obtained from the calculation. This is, of course, assuming that the vibrational frequency has also been scaled accordingly. Thus, as we see an example of in Fig.~\ref{freqRes}, we can use this scaling to obtain sensitivity to different vibrational frequencies. If we have several such systems running concurrently, for example, then we would have achieved sensitivity over a large frequency range, as well as the capacity to distinguish and filter specific desired frequency regions.

\subsection{Experimental prospects}

Here we briefly consider how the above results translate to what can actually be achieved using current superconducting circuit technology. As discussed above, the results that have been presented are invariant under a change of cavity length (fundamental frequency) as long as the other parameters are scaled accordingly. This means, in fact, that the magnitude of signal achieved above is exactly what can be achieved with current technology, since a coupling constant within the neighborhood of $\lambda \sim 0.01 \omega_1$ is achievable in the strong and ultra-strong coupling regimes \cite{Wallraff}. With circuit QED systems one typically has a fundamental frequency   of the order of the GHz's. On the other hand, given Figs.~\ref{corr} and~\ref{freqRes} we see that (somewhat surprisingly) the peak sensitivity of our proposal occurs at $\gamma \sim 10^{-4} \omega_1$. Thus within a cavity QED setup we can expect to be most sensitive to frequencies in the range $\gamma \sim 10^{5}$Hz. Remarkably, this is, for example, on the edge of the frequency range expected from gravitational radiation. 

Given the current state of the art in superconductor technology one can in fact obtain significantly higher coupling constants than what we have considered above. Interestingly, however, we find that this does not significantly increase the seismograph sensitivity as compared with Fig.~\ref{corr}. This is because while the steady state entanglement indeed increases (scaling as $\lambda^2$), the valleys seen in Fig.~\ref{valleys} (which are fundamental to our proposed sensitivity) also become significantly wider and less sharp, meaning that a larger perturbation is required to displace the system out of the valley that it was prepared in. These two changes work against each other in such a way that they largely cancel out. Still, some improvement can be achieved with increased $\lambda$, approaching a signal magnitude near to $2\braket{\op q_1 \op p_2} \approx 6\times 10^{-3}$.  

 This realization may actually be employed to further tune one's apparatus based on experimental restrictions and the strength of vibration one is searching for, such that even typical optical couplings as small as $\lambda =(10^{-6}\omega_1$--$10^{-5}\omega_1)$ may still be useful. By turning $\lambda$ down, the valleys shown in Fig. \ref{valleys} get sharper and thiner (increasing the sensitivity to very weak vibrations), while at the same time reducing the steady state (out of valley) height of the plot (decreasing the maximum response we can obtain). For stronger vibrations having a larger $\lambda$ is preferable since the maximum response (achieved by exiting the valley) is increased. However for very weak vibrations (such that the maximum response is not achieved for larger $\lambda$) it would actually be preferable to have a weaker coupling; a larger response would be observed in this case (assuming that $\lambda$ is not so low as to make the maximum response too small).

\section{Conclusions} 
The quantum seismograph scheme consists of successively sending pairs of atoms in their ground state transversely through an optical cavity. As was known, this will drive the cavity field to a metastable fixed point. 
Here, we found that, surprisingly, the parameters of the farming protocol can be tailored so that the resulting metastable fixed point is highly sensitive to variations of external parameters, such as the cavity length.
 This in turn affects the correlations acquired by the pairs of atoms, which constitutes a detectable signal even for relatively small perturbations. 
  We are proposing to exploit this sensitivity by utilizing the entanglement farming protocol for high precision measurements of small vibrations: a quantum seismograph. 

The proposed sensor has high sensitivity and a sharp spectral response, which should allow one to tune the seismograph to the detection of particular frequencies while screening out noise. The peak frequency in the spectral response of the seismograph can be tuned by adjusting the parameters of the setup.

 This quantum seismograph proposal could be used to detect any kind of vibrational perturbation. In particular, if the cavity walls are coupled through some elastic force,  the passage of a gravitational wave would induce vibrations on the positions of the walls \cite{Weber}, opening the door to potentially using this construction as a novel approach to gravitational wave detection.   Although our current work is still far from making a concrete proposal, we sketch how this scheme may be  adapted to detect the passage of a gravitational wave in Appendix~\ref{app2}.

The sensor's  settings can be easily adapted to carry out measurements of different parameters of the entanglement farming setup, such as the coupling strength, the atomic gap or the travel time of the pairs of atoms. 

Finally, it should also be very interesting to study the quantum seismograph's behavior when the measured parameters behave quantum mechanically. This could potentially yield a new method for measuring mechanical quantum fluctuations, such as those of positions or distances. For example, the length of the cavity may be uncertain due predominantly to quantum fluctuations---e.g.,~if the cavity mirrors are harmonically bound, ultracold, and put into a state that is nearly pure.

\section*{Acknowledgements}
AK and RBM acknowledge support by the Discovery program of the Natural Sciences and Engineering Research Council of Canada (NSERC). EMM acknowleges support by the 
Banting Fellowship program of NSERC. NCM was supported by the Australian Research Council under grant No.~DE120102204.
\appendix

\section{Adiabaticity for moving cavity mirror}\label{app1}

Let us first consider the case of a cavity with one moving wall. 
It is clear that the adiabatic approximation does not hold in general; shaking the cavity walls produces field excitations \cite{Davies1976,dynCas}.
To rigorously see this we must solve the equations of motion subject to time-dependent boundary conditions, and derive an appropriate Hamiltonian that takes such conditions into account. 
We introduce a mode expansion that satisfies Dirichlet boundary conditions (that the field is zero at the moving boundary $\phi(L(t),t)=0$):
\begin{equation}
\phi(x,t)=\sum_n \sqrt\frac{2}{n\pi}\;\varphi_n(t)\sin[k_n(t)x].
\end{equation}
We can now quantize the field system by expressing the Klein-Gordon action in the variables $\varphi_n(t)$ and deriving the Hamiltonian.
Alternatively, we could derive the form of the dynamics by inserting this ansatz into the Klein-Gordon equation and obtaining the form of the expansion coefficients.  
Either way the result is that the field Hamiltonian can be written in terms of the usual stationary solutions plus corrections that are proportional to the time derivative of the boundary condition:
\begin{align}\label{joreba}
\nonumber H &= \sum_n \frac{\omega_n(t)}{2} \left(\pi_n^2(t) + \varphi^2_n(t)\right) \\
&\nonumber - \dot L(t)\sum_{m,n}\alpha_{nm} \omega_n(t) \pi_n(t)\varphi_m(t) \\
&\nonumber + \frac12 \dot L^2(t)\sum_{nm}\left(\sum_k \alpha_{nk}\alpha_{mk}\omega_k(t)\right) \varphi_n(t)\varphi_m(t) \\
& + \frac{\dot L^2(t)}{L(t)}\sum_{m,n}\beta_{nm} \varphi_n(t)\varphi_m(t)
\end{align}
where the canonical conjugate momentum to  $\varphi_n$ is
\[ \pi_n(t)=\frac{\dot \varphi_n(t)}{\omega_n(t)} + \sum_m \alpha_{nm}\varphi_m(t)\dot L (t)\]
The constants $\alpha_{nm}$ and $\beta_{nm}$ are defined by the coefficients of the Fourier sine series of the coefficients of the action  after including all relevant time dependent terms. 
In particular, we have:
\begin{eqnarray}
\alpha_{mn} &=& \begin{cases} 
-\frac{2 (-1)^{m+n} \sqrt{mn}}{\pi (m^2 - n^2)} & \text{if } m \neq n \\ 
\frac{1}{2 \pi n} & \text{if } m = n \end{cases}\\
\beta_{mn} &=& \begin{cases} 
\frac{2 (-1)^{m+n} \sqrt{mn} (m^2 + n^2)}{\pi (m^2 - n^2)^2} & \text{if } m \neq n \\ 
\frac{n \pi}{6} + \frac{1}{4 \pi n}& \text{if } m = n \end{cases}
\end{eqnarray}

Now notice that the first term in \eqref{joreba} is the usual Hamiltonian for the free Klein-Gordon field, but with time-dependent mode frequencies. 
That is exactly what we would get by making the assumption that the field admits the same mode expansion as in the stationary case. 
However, there are other time dependent terms present that allow for particle creation. Those terms are the responsible for mechanisms such as the dynamical Casimir effect.

The remaining terms come with additional factors of $\dot L$, and hence will be small as long as the motion is sufficiently slow.
In particular, the terms in the second line are parametrically smaller by a factor $\dot L$; hence they will be small if the speed of the wall is small compared to the speed of light.
The third and fourth lines are parametrically smaller still by a factor $\dot L$, and hence will be even smaller.
Hence the adiabatic approximation obtains whenever the motion of the cavity walls is nonrelativistic.

 Under the assumption that the cavity walls' motion is non-relativistic, we will have no particle creation. Hence, we can approximate the field dynamics in the cavity by the time dependent mode expansion described above, thus treating the field under the wall oscillations just like a free field in which we make the frequencies and wave numbers time dependent through $L(t)$.  We call this the `adiabatic approximation',   which will be fulfilled in most realistic cases of cavity wall vibration (seismic waves, sound, motion, etc.).

 \section{Adiabaticity for gravitational waves} \label{app2}
 
We now present some preliminary calculations for the case where changes in cavity length are due to a passing gravitational wave. We show that the dynamics in this case are similar to those of the physical model presented in Appendix~\ref{app1} (which was used in our calculations) while acknowledging that further work would be required to refine this into a practical proposal for gravitational wave detection.

It has recently been claimed that relativistic effects in a Bose-Einstein condensate can be used for detection of gravitational waves~\cite{Sabin:2014ww}.
Here we do not make any claims about the achievable sensitivity of our proposal to gravitational waves or the ability to distinguish between gravitational radiation and conventional vibrations; we merely point out that the techniques presented thus far would apply to the case of a gravitational wave.
We leave more detailed questions of sensitivity and isolation from external noise as a possible avenue for future work.

We assume transverse-traceless gauge in which the metric is of the form
\begin{eqnarray} \label{metric}
ds^2 &=& -dt^2 + (\delta_{ij} + h_{ij}(t)) dx^i dx^j 
\end{eqnarray}
where $h$ is a symmetric matrix with entries $|h_{ij}| \ll 1$; the particular form of $h_{ij}(t)$ will depend on the wave profile, and the polarization of the wave.   
Now suppose that the cavity is oriented along the $x$ axis.
The induced metric on the $t-x$ plane will be
\begin{equation}
ds^2 = -dt^2 + (1 + h(t)) dx^2
\end{equation}
where $h(t) = h_{xx}(t)$.
The gravitational wave causes the masses to transversely oscillate and to accelerate toward each other due to the gravitational attraction of the wave between the test masses.
Note, however, that any transverse motion of the masses does not affect the cavity length to first order in $h(t)$.

Suppose the walls of the cavity are 
 coupled  by  a spring.
We will assume that the spring couples the two walls instantaneously.
Although this is not compatible with causality, it should be a reasonable approximation as long as the time scale for forces to propagate   from one end of the cavity to another   is shorter than the typical time scale of the gravitational wave signal.
This condition is determined by the frequency of the gravitational wave $\omega_{gw}$, the speed of sound in the spring $v_s$ and the cavity length $L$, and will hold as long as $\omega_{gw} \ll v_s / L$.

For simplicity, let us describe the walls of the cavity  as a pair of  masses  $m$  (whose separation vector is along the $x$-axis) coupled by an oscillator of quality factor $Q$ and spring constant $k$  (See \cite{Weber}).
We will take the coordinate difference to be $x_1 - x_0 = L_0 + \delta x(t)$, where $L_0$ is the rest length of each cavity (the initial proper separation of the pair of test masses) and $\delta x$ is assumed to be small.
The potential of the spring is 
\begin{equation}
V = \frac{1}{2} k [ (1 + \tfrac12 h(t)) (x_1 - x_0) - L_0]^2.
\end{equation}
The equation of motion can be expanded to linear order in $h(t)$ and $\delta x(t)$, yielding
\begin{equation}
m  {\delta  \ddot x}(t) = - \frac{\omega}{Q} {\delta  \dot x} - k \delta x(t) - \tfrac12 h(t) k L_0.
\end{equation}
We see that the gravitational wave results in a time-dependent external force applied to the spring. 
This results in a time-dependent proper length separations of
\begin{equation} \label{L}
L(t) = L_0 \left( 1 + \tfrac12 h(t) \right) + \delta x (t) 
\end{equation}
and so by solving the equation of motion for $\delta x(t)$ we find the proper length $L(t)$ as a function of time. 

Just as in the case of the moving mirror, we can solve the Klein-Gordon field theory in a mode expansion with moving boundary conditions, the only difference is the time-dependent metric \eqref{metric}.
The resulting time-dependent Hamiltonian takes precisely the same form as Eq.~\eqref{joreba}, but with $\omega_n(t) = n \pi / L(t)$ with $L$ the time-dependent proper length of the cavity \eqref{L}, and additional factors of $\left(1 + \tfrac{1}{2}h(t) \right)$ multiplying the matrices $\alpha_{mn}$ and $\beta_{mn}$. 
Thus the adiabatic approximation holds whenever $\dot L(t)$ is small relative to $c$; we therefore adopt it in all subsequent discussion.

The cavity here described is a linear detector and so cannot distinguish gravitational radiation from tidal forces and other sources of noise.
This could be overcome by considering two (or even three) cavities at right angles and looking for coincident signals.
Gravity waves typically have a strain amplitude of  $h\sim 10^{-21}$ cm.  For a 1 km cavity, this means the driving force is  $k\times 10^{-18}$.  It will be a major challenge to see if our approach could be engineered to achieve such levels of sensitivity.

\bibliography{references,allrefs}

\end{document}